**Multi-wavelength Study of Energetic Processes during Solar Flares occurrence**

Shirsh Lata Soni[1*], Radhe Shyam. Gupta[1**], Adya Prasad Mishra[2***]

[1] Department of Physics, Govt.P.G. College satna MP,India, 485001
[2] Department of Physics, Awadhesh Pratap Singh University, Rewa, MP India
[*]*sheersh171@gmail.com*
[**]*rsgsatna@gmail.com*
[***]*adyaaps@gmail.com*

**Abstract:**

This paper is an attempt to understand the physical processes occurring in different layers of solar atmosphere during a solar flare. For a complete understanding of the flare we must analyze multi-wavelength datasets, as emission at different wavelength originates from different layers of the solar atmosphere. Also, flares are transient and localized events observed to occur at all longitudes. With these considerations, we have carried out multi-wavelength analysis of two representative flare events. One event occurred close to the centre of the solar disk and other occurred close to the limb. In the former case, we examine the emission from the lower layers of the solar atmosphere. Therefore the chromospheres, transition region and also the photospheric magneto-gram can be analyzed. On the other hand, in near-limb event, coronal features can be clearly studied. In this paper, the first event studied is the M 1.1 class flare from the active region NOAA 10649 located at S10E14 and the second event is the M 1.4 class flare from the active region NOAA 10713 located at S12W90. In both cases, we have observed the excellent multi-wavelength data sets. The observations from multi-instrumental data clearly shows that flares occur in the vicinity of sunspots. These are regions of strong magnetic field with mixed polarity.

## 1. Introduction

The study of multi-wavelength emission during solar flares has enormous potential towards understanding the underlying physical phenomena occurring in the solar atmosphere. It is generally accepted that magnetic reconnection is responsible for the sudden energy release and acceleration of particle in solar exploration (solar flare) (Joshi B., et al. 2011). In solar flare, the deviation of the magnetic field from the potential field is associated with the energy storage and release process. Therefore, researchers studied the temporal and special correspondence between flare occurrence and photospheric magnetic field properties in order to characterize the magnetic non-potentiality (e.g., Hagyard et al. 1984; Hagyard 1990; Abramenko et al. 1991; Wang et al. 2007, Cui et al. 2006 ; Pevtsov et al. 1995; Deng et al. 2001; Moon et al. 2002; Leka & Barnes 2013a, 2013b). Since the highly sheared and highly twisted magnetic fields are usually accompanied with strong current. The other important quantity used to study and monitor the evolution of magnetic non-potentiality in flare is electric current. On other hand, based on multi-wavelength study, a standard model called the CSHKP model was developed (Carmichael 1964; Sturrock 1966; Hirayama 1974; Kopp & Pneuman 1976), which can explain the flare features of the separating two ribbons observed by H-alpha/UV and the expanding soft X-ray (SXR) loops.

In this present study we analyse the physical process of energy release during solar flare. These solar flare occurred in different longitude on solar disc so here it is important to study behaviour of the solar flare emission with different angle. For that we have chosen two flares, one of them occurred near the centre of the solar disk at the south-east region, with the heliographic co-ordinates S10E14 from Active region NOAA 10649 on 17 July 2004 and other of is limb event which occurred on 29 December 2004 in the active region NOA 10713, located close to the limb of the Sun at the south-west region, with the heliographic co-coordinates S12W90. Both the flares are nearly same class of intensity M-class (REF: GOES intensity classification). With the help of various imaging instruments on-board in different satellites with multiple energy bands and time resolutions, we are trying to analyse the energy release process during flares.

The study of solar flare is exciting and challenging. The observations of solar flares provide us a wealth of knowledge about the basic plasma processes (such as magnetic reconnection) and behaviour of magnetized plasma in high temperature environments. Further, the understanding the behaviour of solar flares enable us to study similar process in other stars and astrophysical systems. This required through multi-wavelength observation of the Sun with high temperature and special resolutions. Technological developments have provide us with the capability to look further into fact that have been invisible in the past. Future scientific research would lead to better understanding of the Sun-Earth system.

## 2. Data Sources

The present analysis is based on the data taken from the following instruments: magnetograms and intensity-grams observed by SOHO/Michelson Doppler Imager (MDI); X-ray measurements from GOES and RHESSI; UV and EUV images from TRACE.

The Transition Region and Coronal Explorer (TRACE) obtained 1 arcsec resolution, multithermal images of the solar atmosphere with a cadence and continuity of coverage commensurate with the variability and lifetimes of primary solar targets such as X-rays bright points, active regions, filaments and flares. TRACE combined with UV and EUV multilayers and different wavelength filters (2500 Å,1700 Å,1600 Å,1550 Å,1216 Å,284 Å,195 Å,171 Å). For the present study we have used the 1600 Å filter images which image chromospheric features for UV region and 195 Å filter images correspond to the coronal region of the solar atmosphere and the filter is sensitive to a temperature of about 1.5 million K. SOHO/ MDI measures line-of-sight motion (Dopplergrams), magnetic field (magnetograms), and brightness images in full disk at several resolutions (4 arcsec to very low resolution) and a fixed selected region in higher resolution (1.2 arcsec). GOES X-ray sensors observe the Sun continuously in two broadband soft X-ray channel (1-8 Å) and the harder channel (0.5-4 Å) which provides the information of coronal plasma. From GOES X-ray data we can compute the effective colour temperature and emission measure (EM) for the solar flare with 3-sec time resolution. Reuven Ramaty High Energy Solar Spectroscopic Imager (RHESSI) is designed to investigate particle acceleration and energy release in solar flares through imaging and spectroscopy of hard X-ray and gamma-rays. It images solar flares in energetic photons from soft X-rays (~3 keV) to gamma rays (up to ~20 MeV) and to provide high resolution spectroscopy up to gamma-ray energies of ~20 MeV. Its rotating modulated collimators allows angular resolution down to 2.3 arcsec. The data analysis in this paper has been performed using interactive data language (IDL) programming and Solarsoft.

## 3. Multi-wavelength observations of two M-class solar flares

The standard flare model, also known as CSHKP model, recognizes that the evolution of flare loops and ribbon can be understood as a consequence of the relaxation of magnetic field lines stretched by the ejection of plasma (Carmichael, 1964, Hirayama T., 1966). The magnetic reconnection has been identified as the key process which releases sufficient magnetic energy on short time scales to account for the radiative and kinetic energies observed during an eruptive event (Priest E.SR., 2002). A solar flare is a multi-wavelength phenomenon. Therefore in order to have a complete understanding of its temporal evolution we need to look at the time profiles observed at different wavelengths. However, it has been observed that there could be subtle activities at the flare location before its onset. In the following, we discuss different aspects of flare evolution.

### 3.1 M 1.1 class flares on 17 july2004

The MDI instrument on the SOHO provides almost continuous observations of the Sun and records the intensity-grams and magneto-grams. The first event selected for the study in this paper occurred on 17 July 2004. The flare was observed in the active region NOAA 10649, located near the centre of the solar disk at the south-east region, with the heliographic co-ordinates S10E14. The intensity-grams of the active region NOAA 10649 is shown in Figure 1 (a). The image clearly shows groups of leading as well as following sunspots (indicated by arrows). The image represents the magnetic field of the active region in black and white colours which correspond to the region of south and north polarity, respectively. Although the leading sunspot is larger in size, it has a simplified magnetic structure. However, the following sunspot has a complex magnetic structure with mixed polarity.

Figure 2 shows the GOES soft X-ray  light curve obtained in two different channels (wavelengths 1.0 -8.0  Å and 0.5-4.0 Å). The wavelength 1.0 – 8.0 Å corresponds to energy 12.5-1.5 keV and 0.5-4.0 Å corresponds to energy 24.0 -3.0 keV. According to the GOES profile, the event took place between 22:54 UT and 23:20UT, with maximum emission at 23:00 UT. It is evident that, there is a gradual rise during the initial phase of the flare. It is interesting to note that there is no significant decrease in the soft X-ray flux for ~10 minutes after the peak at 23:00 UT. This feature of the soft X-ray profile is more noticeable in 1.0 -8.0 Å light curve, in which the flux is almost sustained between 23:00 UT and 23:10 UT. Then this event continues to decline gradually after ~23:10 UT. From the GOES flux data, it is recorded as M1.1 class soft X-ray flare based on the peak intensity of emission.

The flare was also observed by the Reuven Ramaty High Energy Solar Spectroscopic Imager (RHESSI) spacecraft. The RHESSI X-ray light curves are obtained for energy bands of 6-12 keV, 12-25 keV and 25-50 keV, as seen in Figure 2.3. Similar to the GOES profile, the rise of the initial phase is observed at ~22:54 UT in both 6-12 keV and 12-25 keV energy bands.

The decay phase begins at 23:10 UT in the 6-12 keV and 12-25 keV light curves.

However, the hard X-ray profile is impulsive in nature and shows a maximum intensity at ~22:58 UT. The difference of the timing in the observation of the maximum phase between hard X-ray and soft X-ray(~2 minutes) is noteworthy.

The flare observations were also made by the transition region and Coronal Explorer (TRACE) satellite. TRACE observations of the flare were obtained in EUV and UV spectral range, using the filters of wavelengths 195 and 1600 respectively.

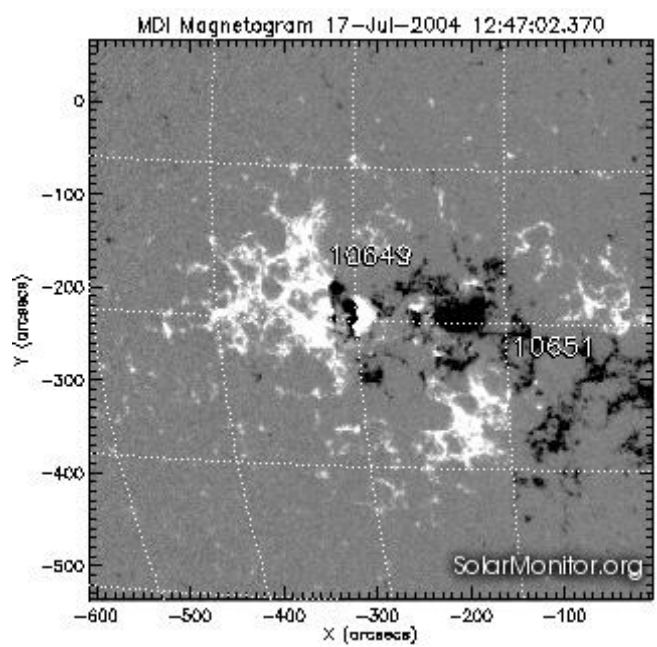

Figure: 1 SOHO/MDI intensity gram and magneto gram observed on 17 July 2004. The images show the active region NOAA 10649.

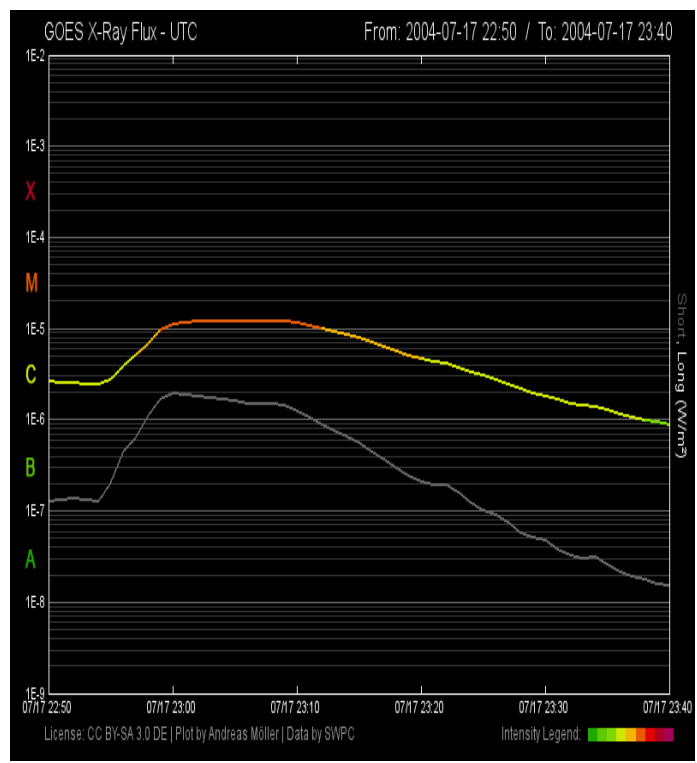

Figure: 2 Temporal evolutions of GOES soft X-ray for the M 1.1 flare on 17 July 2004. The coloured line shows the plot in the 1.0- 8.0 Å (12.5- 1.5keV) channel and the solid grey line represents the plot in the 0.5 – 4.0 Å (24.0 – 3.0 keV) channel.

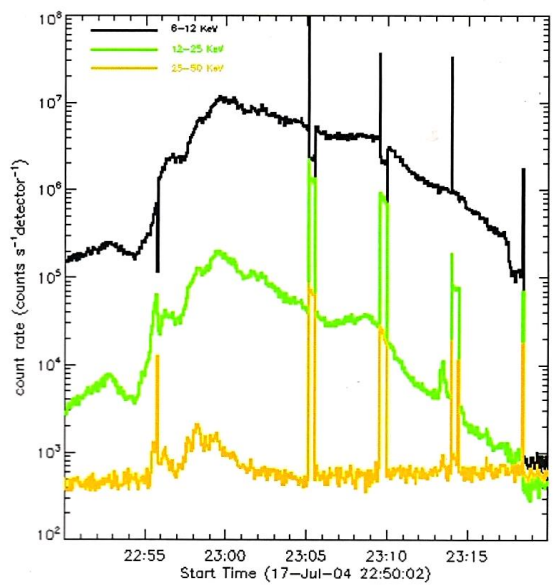

Figure: 3 Temporal evolution of X-ray plot from RHESSI. The RHESSI time profiles are obtained for energy bands of 6-12 keV, 12-25 keV and 25-50 keV. The sharp changes in the time profile are due to the instrument artefact corresponding to change in the attenuator state.

The TRACE 195 Å images correspond to the coronal region of the solar atmosphere and the filter is sensitive to a temperature of about 1.5 million K. The observations of the M1.1 class flare in TRACE 195 Å have been recorded completely from the initial phase of the flare to the final phase. Figure 4 shows the series of flare images by TRACE 195 Å. The images recorded during the initial phase (~22:55:20 UT) clearly shows that the emission originates from two regions in the flaring location. For convenience let us call them as the eastern brightening and western brightening, respectively. These two bright regions are well separated from one another and their structures are clearly observed. The magnetic loops connecting the regions are also visible in the images.

As the flare evolve, the structure of the two bright regions changes, as seen in the Figure 4 at 23:00:32 UT. A little while later, these bright regions are no longer well resolved. They are connected to one another, another region appears as one bright structure shaped like a V, as seen at 23:02:10 UT. This connection between the eastern and western brightening is due to the release of high energy by the process of magnetic reconnection. As the energy is released, the plasma is heated to high temperatures and the separation between the two regions is filled by the hot plasma.

Further as the plasma cools, the two regions become well resolved again, as seen at 23:02:21 UT. Yet again the energy is released and hot plasma fills the separation and cools at a later time, the series of EUV images suggest that the process of pumping of plasma is repeated a few times until 23:10:51 UT. We interpret this process as the distinct events of energy release during the flare evolution.

This process of episodic energy release is clearly reflected in the GOES soft X-ray light curve in the form of prolonged maximum phase. As discussed earlier, the maximum phase is observed to be sustained from 23:00 UT to 23:10 UT and the corresponding analysis of TRACE 195Å between these periods provides a consistent picture the bright fuzzy structure observed thin EUV TRACE images corresponds to plasma of very high temperature. It is also observed that at high temperature during release of energy, the plasma spreads out and is observed in a larger volume. As the temperature reduces, the structures are resolved and the hot plasma now occupies lesser volume. Throughout the flare, the eastern brightening evolves and significant change in its structure is clearly seen, due to the energy release process. Therefore, from the comparison of the X- ray light curve to the EUV images, the prolonged maximum phase of the flare can be well understood.

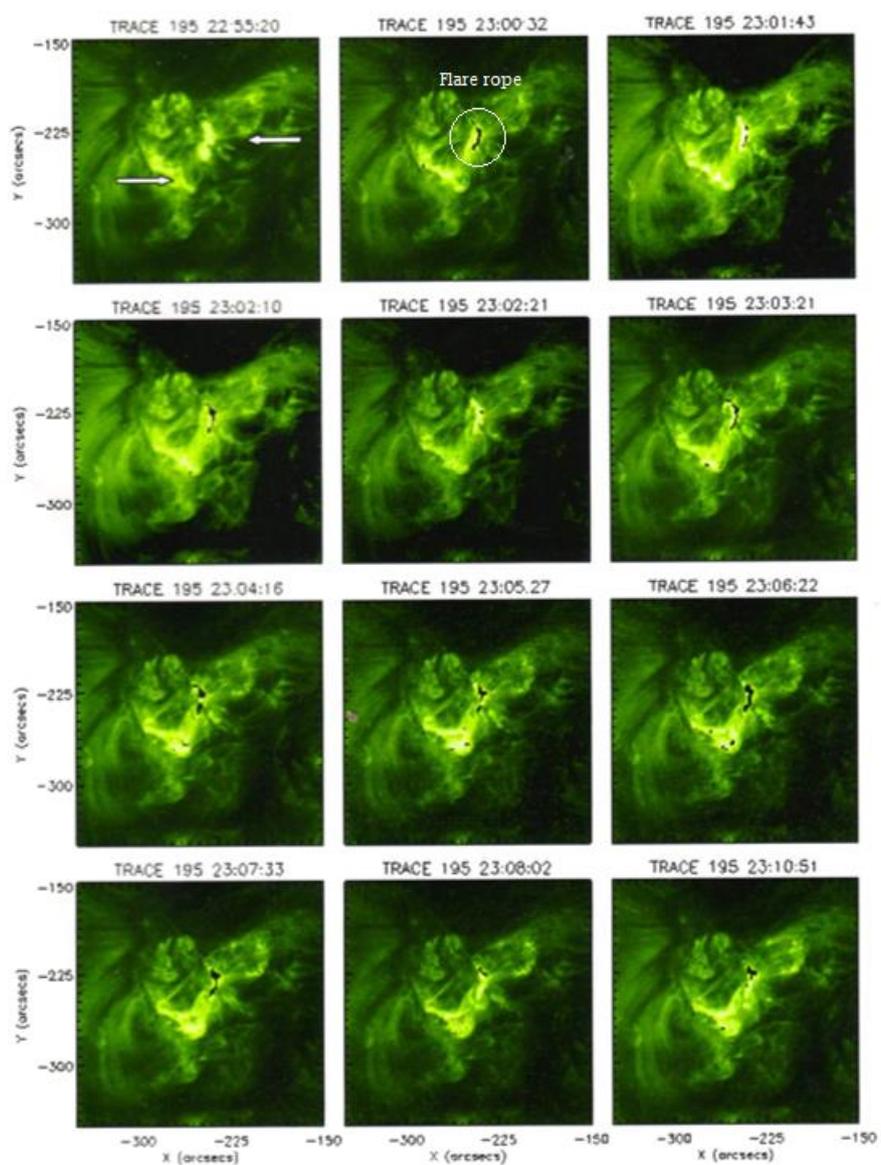

Figure: 4 Sequential TRACE 195 Å images of active region NOAA 10649 during the flare. The images are recorded from the initial phase (22:50:20 UT) to the decay phase  (23:10:51 UT) of the flare.

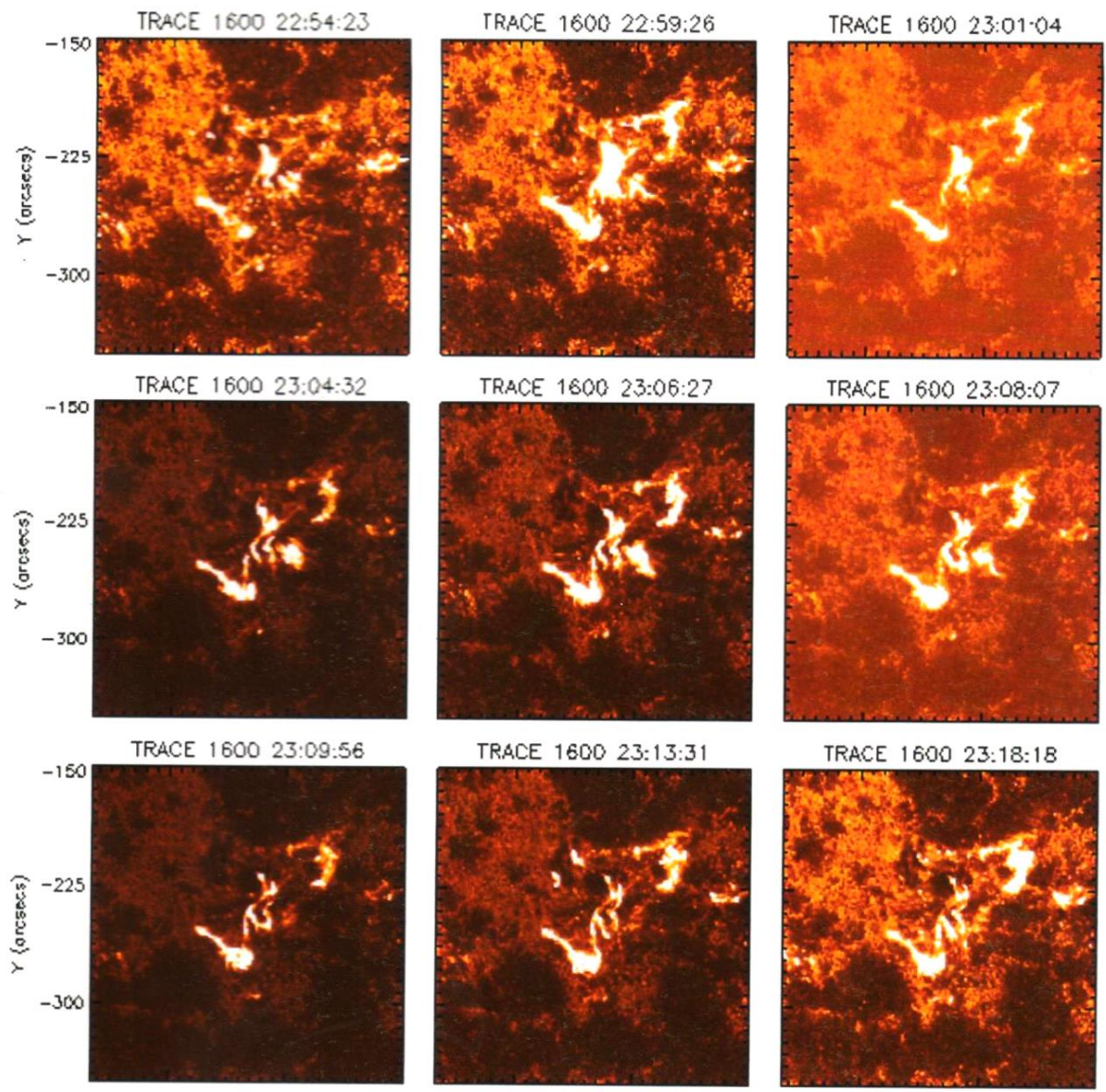

Figure: 5 Sequential TRACE 1600 Å Images of active region NOAA 10649 during the flare. The images are recorded from the initial phase (22:54:23 UT) to the late decay phase (23:18:18 UT) of the flare.

The TRACE 1600 Å image corresponds to the chromospheric and transition region of the solar atmosphere and is sensitive to temperature of the range 5000-10000K. The observations of the M1.1 class flare in TRACE 1600 Å has also been recorded from the initial phase to the decay phase. The series of images of the flare from the TRACE 1600 Å is shown in Figure 5. The bright regions, as seen in 195 filter images are also observed here. This provides evidence that during the flare there is link in the processes occurring in the different layers of the solar atmosphere extending from the corona to the chromospheres. Similar to the TRACE 195 Å images, two well resolved bright regions are observed. Also several ribbon-like bright regions are electrons in the lower layers of the Sun. The electrons present in the solar corona are accelerated distinct changes in the active region are observed until 23:13:31UT. We observe rather small changes during the decay phase of the flare.

## 3.2 M 1.4 class flare on 29 December2004

The second flare event selected for the study in this work occurred on 29 December 2004. The flare was observed in the active region NOAA 10713, located close to the limb of the Sun at the south-west region, with the heliographic co-coordinates S12W90. The present analysis is based on the data taken from the following instruments: magnetogrms and intensity grams observed by SOHO/Michelson Doppler Imager; X-ray measurements from GOES and RHESSI; EUV image.

In Figure 6 we present the magneto-gram of the active region NOAA 10713 on 27 December 2004 that represents the magnetic field strength and polarity in the Sun's photosphere. On the day of the occurrence of the flare the region of interest was located very close to the solar limb. In such situations, we hardly resolve any detailed structures due to the projection effects. Therefore, we have selected the magneto-gram of the active region two days prior to the event. The image represents the magnetic field of the active region in black and white colours which correspond to the region of south and north polarity, respectively.

Figure 7 shows the GOES X-ray flux light curve obtained in two different channels with wavelengths 1.0 -8.0 and 0.5 -4.0.accroding to the GOES light curve, the event took place between 19:12 UT and 19:40 UT with maximum emission at 19:20 UT. The initial phase of the flare, there is a steep rise from ~19:12 UT. However, at 19:15 UT there is a change in the trend of the light curve. Again at 19:18 UT there is a sharp and fast rise. Thus, we note that, the evolution of the flare takes place in two stages. After peak at ~19:20 UT, the soft X-ray flux begins to decay. We observe a faster decay until 19:27 UT. Additionally, we observe gradual decline in the soft X-ray flux. Hence, the flare also decays in two stages. From the GOES flux data, it is recorded as M 1.4 class soft X-ray flare.

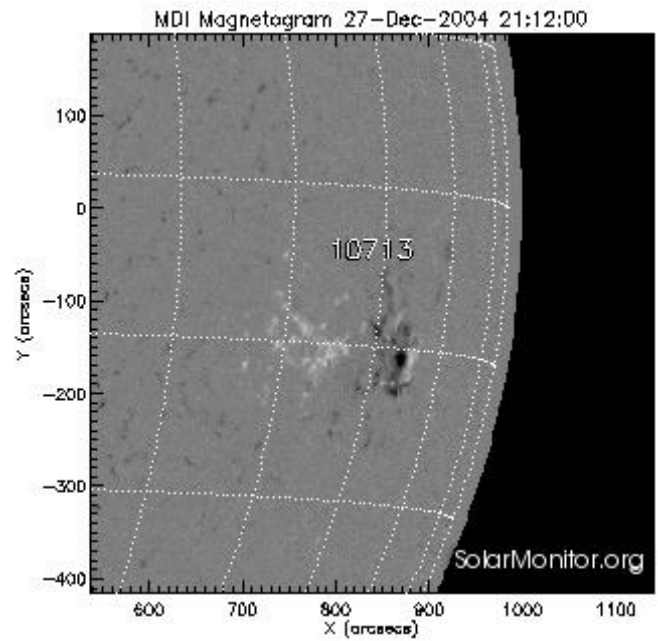

Figure: 6 SOHO/MDI magnetogram observed on 27 December 2004, two days prior to the event studied. The image shows the active region NOAA 10713.

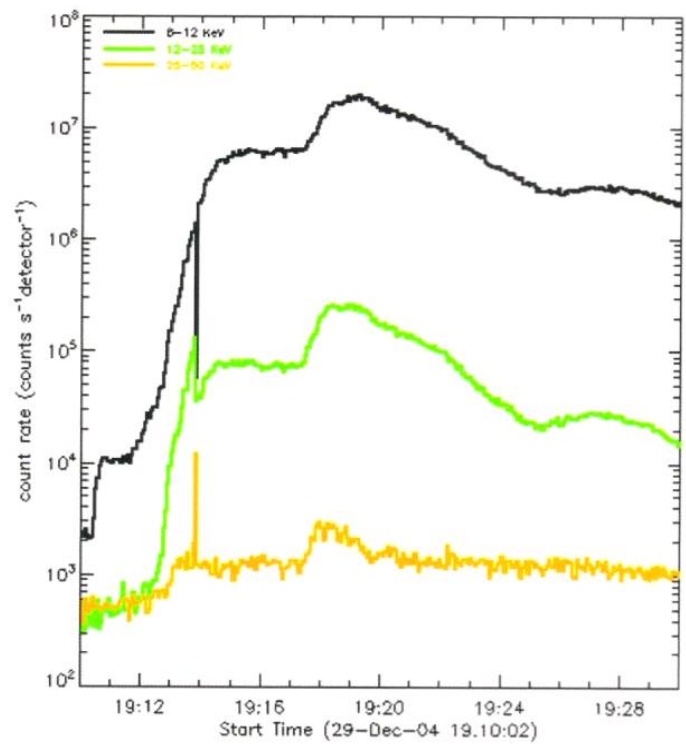

Figure: 7 Temporal evolution of X-rays plot from RHESSI. The RHESSI. The RHESSI time profiles are obtained for energy bands of 6 -12 keV, 12 -25 keV and 25-50 keV . The sharp changes in the time profile are due to the instrument artefact corresponding to change in the attenuator state.

The flare was also observed by the RHESSI space craft. The RHESSI X-ray light curves are obtained for energy bands of 6-12 keV, 12-25 keV and 25-50 keV, as seen in Figure 7. It was also found that emission observed above than 50 keV energy band. The variations in the RHESSI light curves are very similar to the GOES profile (Figure 8). Here also, we clearly see the two-stage evolution of the flare in the rise phase. The overall maximum of the flare coincides at 19:19 UT, in all three energy channels. The difference of the timing in the observation of the maximum phase between hard X-ray and soft X-ray is ~1 minute. From the RHESSI measurement, the two phases declination are evident in the decay phase of the flare as well, consistent with the GOES findings. The decay phase begins ~19:21 UT and at~19:26 UT the trend changes in the light curve.

The flare observations were also made by the TRACE satellite. TRACE observations for the flare were obtained in the UV range, using the filter of wavelength 1600. The observation of the M 1.4 class flare using this filter is recorded from the initial phase to the decay phase. After the onset of the flare, a loop is observed in the southern part of the active region and it is the pre-flare loop. This is clearly seen from the Figure 9 at 19:12:38 UT. Also, there is a small bright region observed to be present in the northern part of the active region and it could be possibly unresolved loop structure.

As the flare evolves, the pre-existing loop bright regions changes, as seen in the Figure 10 at 19:13:47 UT. During the progression of the flare to be the maximum phase 19:15:57 UT, this loop structure gets disrupted. This corresponds to the change in trend recorded in the GOES light curve. During the maximum phase, at ~19:20:08 UT, intense brightening is observed from both the northern and southern regions of the activity site. We find that the northern part of the active region sustains its configuration, while the southern loop system undergoes complete structural change. After 19:25:34 UT, the region does not show significant activities. This clearly indicates the second gradual stage of the decay phase during which hot plasma cools down slowly.

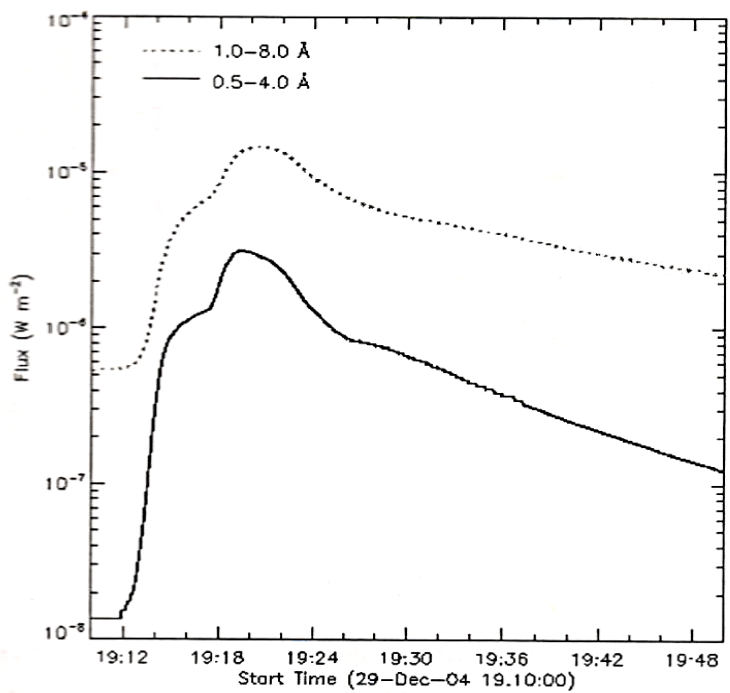

Figure: 8 Temporal evolution of GOES soft X-rays for the M 1.4 flare on 29 December 2004. The dotted line shows the plot in the 1.0 − 8.0 (12.5 − 1.5 keV) channel while the solid line represents the plot in the 0.5 − 4.0 (24 − 30 keV) channel.

The emphasis is given on observations taken by the Reuven Ramaty High Energy Solar Spectroscopic Imager (RHESSI) on the X-ray emission originating from the coronal loops. (Lin et al. 2002; Joshi B., 2012) The RHESSI X-ray image for the M1.4 class limb flare is obtained at the 6-2 keV energy band. The loop top observed at the positions x=957 arcsec, y=-218 arcsec on the solar grid. By comparing the partial positions of RHESSI X-ray sources with the TRACE1600 image, We find that X-ray emission at this energy band is originating from the coronal loops, which gets disrupted during the flare.

We further analyse time evolution of 10-15 keV X-ray source during the flare from its onset to decay phase in Figure 11. We find that the X-ray source is located above the solar limb throughout the event. This suggests that X-ray source moves upward with the flare evolution. This indicates that the energy release site, that is the region of magnetic reconnection, shifts to higher altitude in the corona during the flare.

Figure: 9 Sequential TRACE 1600 images of active region NOAA 10713 during the flare. The images are recorded from the initial phase (19:12:38 UT) to the decay phase (19:27:36 UT) of the flare.

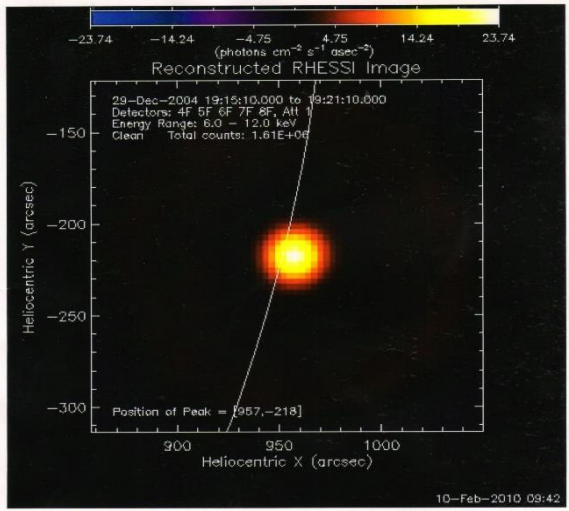

Figure: 10 Reconstructed RHESSI Hard X- ray image of the solar flare in the energy rangy range 6-12 keV.

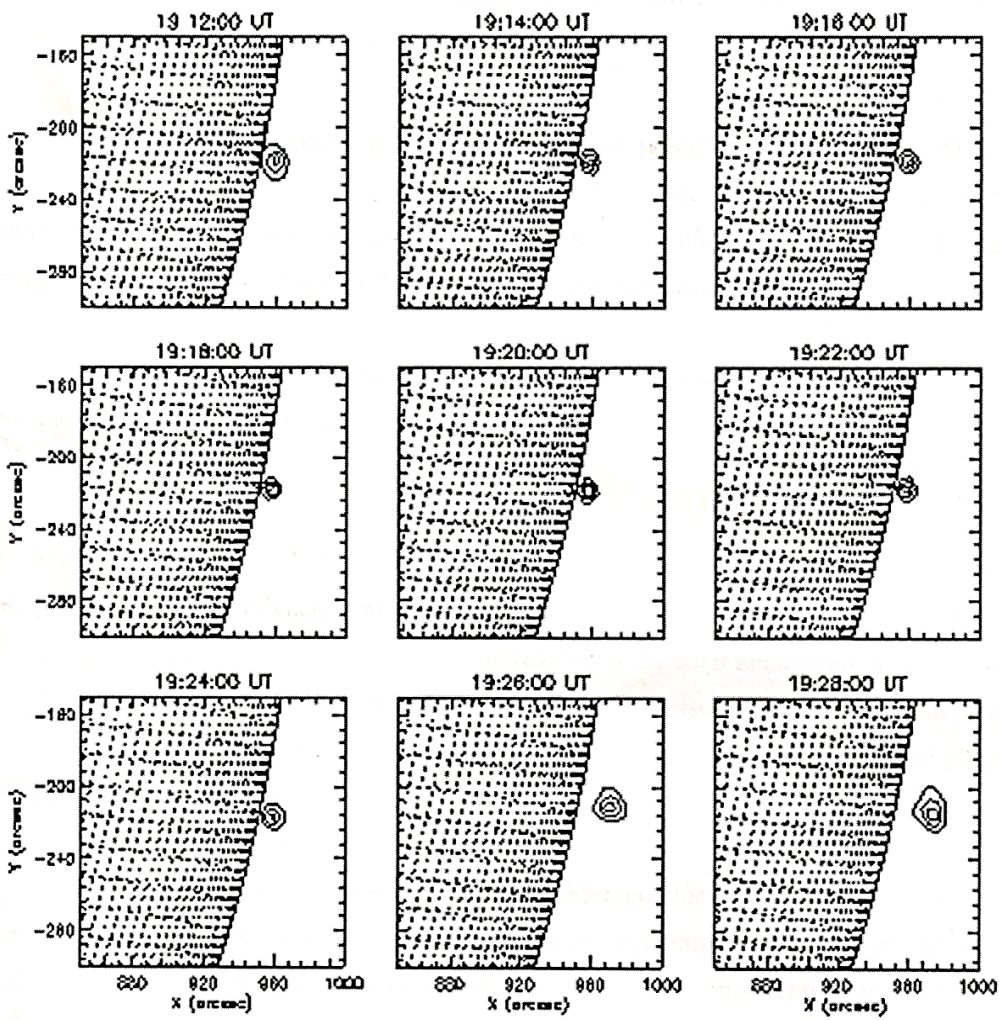

Figure: 11 Large scale contraction of coronal loops observed for ~16 minutes in an M1.4 flare that occurred in active region NOAA 10713 on December 29, 2004 at a location S12W90. The contraction of coronal loops can be readily seen in TRACE EUV images at 1600 ˚A (previous image) and RHESSI X-ray images at 10-15 keV.

### 4. Summary and conclusion

This presented work focuses on understanding the energetic processes in solar flares. We have selected two M class solar flares, one occurring close to the centre of the Sun and other close to the limb. Excellent set of multi-wavelength data enables us to understand signature of energy release process in different regions of the solar atmosphere in different wavelengths.

A flare is a magnetic phenomenon. Data from intensity-grams and magneto-grams clearly shows that flares occur in the vicinity of sunspots. These are regions of strong magnetic field with mixed polarity. Flares are accompanied by emission at different wavelengths. The flares studied in this paper provide a scenario of energy release process in hard X-ray, soft X-ray, extreme ultraviolet and ultraviolet. Thus a flaring process involves all regions of the solar atmosphere. Observations of regions of brightening in ultraviolet, extreme-ultraviolet and also enhancement in soft X-ray flux suggest that plasma is heated to different temperatures. This provides evidence that plasma in flaring region is multi-temperature.

High energy hard X-ray emission was also observed during flares. The hard x-ray profiles are impulsive. This suggests explosive release of huge amount of energy in very short time scale. Theoreticians believe that such violent release of energy on short time scale can only be explained by the magnetic reconnection process. Magnetic reconnection is defined as he breaking and topological rearrangement of oppositely directed magnetic field lines in a plasma. In this process, magnetic field energy is converted to plasma, kinetic energy and thermal energy. A simple schematic diagram of the magnetic reconnection process is shown in Figure 12. In our analysis, we detected an X-ray coronal source in the near-limb event. This is an evidentiary support for the process of magnetic reconnection in the solar corona. Due to impulsive energy release in the corona, particles (mostly electrons) are accelerated. Also the lower show emission in longer wavelengths of UV and EUV bands. In our flares, we find that the magnetic loop structure of the loops are observed along with the formation of new loops due to coronal energy release.

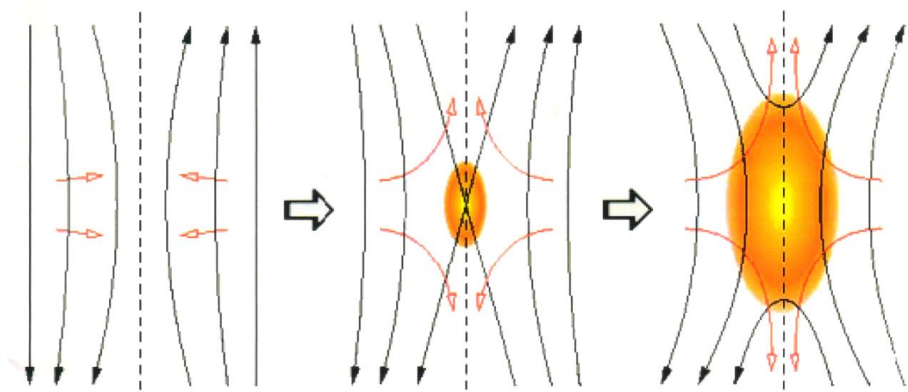

Figure: 12 Schematic representation of magnetic reconnection process.

In the limb event, we found that the X-ray source moves upward in the corona indicating that the energy release site shifts to higher altitudes. Therefore, we understand that the primary energy release during a flare takes place in the corona. These observations can be understood from the standard flare model shown in Figure 13. In this simplified picture, energy release takes place above the magnetic field loops that extend from the photosphere into the corona. We can see that the foot points of the flare loops are anchored in the chromospheres, in regions of opposite magnetic polarity. Stored magnetic energy is released above the top of the loops due to magnetic reconnection. The process of energy release causes acceleration of particles (mostly electrons). The electrons are accelerated to high speed, generating impulsive hard X-ray emission. Some of the non-thermal electron are channelled down and penetrate the chromospheres at high speed heating the plasma and thus intense brightening is observed in hard X-ray, UV and H-alpha. When beams of accelerated protons enter the dense, lower atmosphere, they may produce nuclear reactions resulting in gamma-rays in some exceptionally high energetic flares. Materials in the chromospheres are heated quickly and rise into the loops which results in a slow gradual increase in soft X-ray radiation.

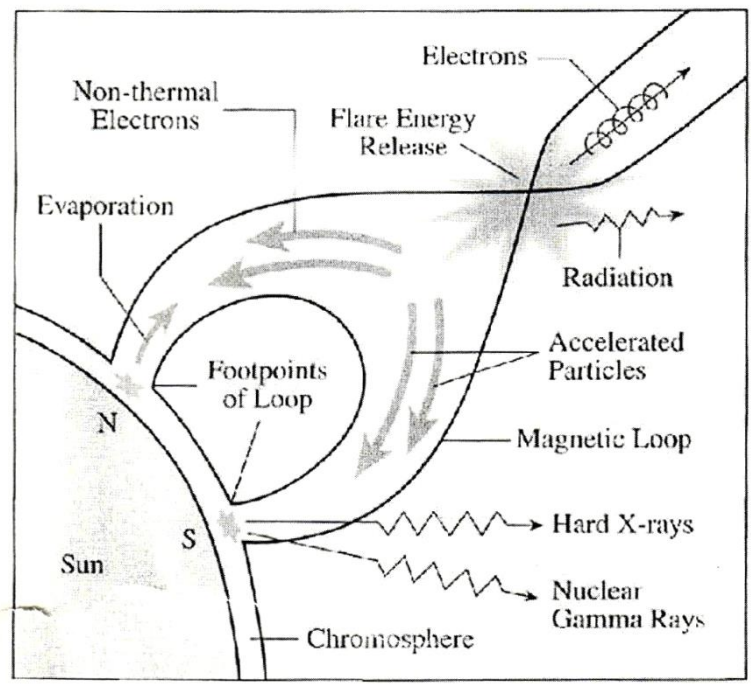

Figure: 13 Standard flare Model.

In the disk event presented in this paper, we clearly observed brightening of the lower and denser atmospheric layers, that is the chromospheric and transition region, in the form of ribbon-like structures in UV images.

From the above study we can conclude that the multi-wavelength and multi instrumental analysis have improved our understanding of various physical processes during solar flare occurring in different layers of solar atmosphere. The standard flare models broadly recognised these physical processes as the consequence of large scale reconnection of magnetic field lines in the corona. However, the modern and advancement in the observational equipment capabilities has led to several new insights of the flare evolution that deviate from the standard models. As we know that flares have been observed from last more than 150 years and many investigators studied them from last 3-4 decades but still we are far away from full understand of evolution of flares. We have yet to understand various basic elements cause to pre flare magnetic configuration, energy release process and site, triggering mechanism, conversion of magnetic energy to heat and kinetic energy, particle acceleration etc. these unsolved issues pose challenges for future researchers.

**Acknowledgements** Authors are grateful to CME catalogue from SOHO/LASCO observation maintained and generated by CDAW data Centre by NASA. We thank Prof. Bhuwan Joshi,USO, Physical Research Laboratory, Ahmedabad, and Pro. Hari Om Vats, scientist, Physical Research Laboratory, Ahmedabad India for his great support to us. Thanks are also due to an anonymous referee for helpful suggestions.